\documentstyle[preprint,prd,aps]{revtex}
\newcommand{\be}{\begin{equation}}
\newcommand{\ee}{\end{equation}}
\newcommand{\bea}{\begin{eqnarray}}
\newcommand{\eea}{\end{eqnarray}}

\newcommand{\sptwo}{1.4}
\newcommand{\doublespace}{\edef\baselinestretch{\sptwo}\Large\normalsize}

\newcommand{\newsection}[1]{\section{#1}
\setcounter{equation}{0}}

\newcounter{newapp}
\begin{document}
\begin{titlepage}

\setcounter{page}{1}

\vspace{1.5in}

\title{Nonlinear Realization of Supersymmetry and Superconformal Symmetry}
\author{T.E. Clark\footnote{e-mail address: clark@physics.purdue.edu} and S.T. 
Love\footnote{e-mail address: love@physics.purdue.edu} }
\address{\it Department of Physics, 
Purdue University,
West Lafayette, IN 47907-2036}
\maketitle
\begin{abstract}
Nonlinear realizations describing the spontaneous breakown of supersymmetry and $R$ symmetry are constructed using the Goldstino and $R$ axion fields. The associated $R$ current, supersymmetry current and energy-momentum tensor are shown to be related under the nonlinear supersymmetry transformations. Nonlinear realizations of the superconformal algebra carried by these degrees of freedom are also displayed. The divergences of the $R$ and dilatation currents are related to the divergence of the superconformal currents through nonlinear supersymmetry transformations which in turn relates the explicit breakings of these symmetries.   
\end{abstract}

\newpage
\doublespace
\end{titlepage}

\newsection{Introduction}

Effective Lagrangians based on nonlinear realizations of spontaneously broken symmetries provide an extremely useful, model independent, way of ensapsulating the dynamical constraints mandated by the symmetry breakdown\cite{SW}-\cite{CCWZ}. Such techniques have been successfully applied to a wide range of physical problems most notably in the form of nonlinear sigma models\cite{EL}. In this paper, we construct nonlinear realizations of spontaneously broken supersymmetry (SUSY) and $R$ symmetry. If supersymmetry is to be realized in nature, it must be as a broken symmetry. The breaking mechanism which maintains the perponderance of the dynamical constraints of the symmetry and hence is theoretically most attractive, is a spontaneous one. 

Thus we envision some underlying theory in which both the supersymmetry and the $R$ symmetry are spontaneously broken. The specific dynamics responsible for the symmetry breakings is left unspecified. As such, we allow for the possibility that the dynamics producing the SUSY breaking has a completely different origin than that producing the $R$ symmetry breaking and moreover that the scales at which the symmetry breakings occur could be completely independent. 

The relevant effective Lagrangian describing the low energy degrees of freedom contains the Nambu-Goldstone fermion of spontaneously broken supersymmetry\cite{AV}-\cite{FI}, the Goldstino, and the (pseudo-) Nambu-Goldstone boson of spontaneously broken $R$ symmetry, the  $R$-axion. Moreover, if the spontaneously broken supersymmetry is gauged, the erstwhile Goldstino degrees of freedom  are absorbed to become the longitudinal (spin 1/2) modes of the spin 3/2 gravitino via the super-Higgs mechanism. As such the dynamics of those modes are given by that of the Goldstino. In the next section, we construct the nonlinear realization of SUSY in terms of these Nambu-Goldstone modes leading to actions invariant under nonlinear SUSY, but possibly containing some soft explicit $R$ symmetry breaking in addition to its spontaneous breaking.  Next, the $R$ current, supersymmetry currents and energy-momentum tensor are obtained and shown to be related under nonlinear SUSY transformations. Finally, we construct the nonlinear realization of the superconformal algebra using these Nambu-Goldstone fields and form the superconformal currents and dilatation currents. We explicitly display how the divergence of the $R$ current and divergence of the dilatation current are related to the divergence of the superconformal current through the nonlinear SUSY transformation thus relating the various $R$, scale and superconformal explicit symmetry  breaking terms in the effective Lagrangian.

\newpage

\newsection{Nonlinear realizations of SUSY and $R$ symmetry: invariant actions}

A method\cite{CCWZ} for constructing nonlinear realizations of spontaneously broken internal symmetries employs the construction of the coset group element whose coset space coordinates are the Nambu-Goldstone bosons and then extracting the changes in these coordinates under group multiplication. This procedure requires a slight modification for spontaneously broken spacetime symmetries\cite{V}-\cite{O}. This follows since motion in the coset space is accompanied by motion in spacetime. Since the supersymmetry generators, $Q_\alpha, \bar{Q}_{\dot\alpha}$, and the $R$-symmetry charge together satisfy the supersymmetry algebra\cite{W}:
\bea
&&\{ Q_\alpha , \bar{Q}_{\dot\alpha} \} = 2\sigma^\mu_{\alpha \dot\alpha} P_\mu ~~;~~\{ Q_\alpha , Q_\beta \}=\{ \bar{Q}_{\dot\alpha} , \bar{Q}_{\dot\beta} \}=0 \cr
&&[P^\mu, Q_\alpha]=[P^\mu, \bar{Q}_{\dot\alpha}]=[P^\mu, R]=0 \cr
&&[R, Q_{\alpha} ] = Q_{\alpha}~~;~~[R, \bar{Q}_{\dot\alpha} ] = -\bar{Q}_{\dot\alpha} \, ,
\label{SSA}
\eea
where $P^\mu$ are the spacetime translation generators,
it follows that both supersymmetry and $R$ symmetry constitute spacetime symmetries. Consequently, to construct  nonlinear realizations of spontaneously broken supersymmetry and $R$ symmetry,  one needs to include the product of the unbroken translation group element along with the coset group elements. 

Such a product of translation and coset group elements is given by
\be
\Omega(x,\lambda,\bar\lambda, a) = e^{-ix_\mu P^\mu} e^{i(\lambda^\alpha(x) Q_\alpha +\bar\lambda_{\dot\alpha}(x) \bar{Q}^{\dot\alpha}) }e^{i a(x) R} \equiv \Omega (x)
\, .
\ee
Here $\lambda^\alpha, \bar\lambda_{\dot\alpha}$ are the 2-component Weyl spinor Goldstino fields of the spontaneously broken supersymmetry and $a$ is the $R$-axion field, the (pseudo-) Nambu-Goldstone boson of the spontaneously broken $R$ symmetry. They act as the coordinates of coset space corresponding to symmetry pattern: Poincar\'e $\times$ Supersymmetry $\times$ R $\rightarrow$ Poincar\'e. Next define the group element 
\be
g(0, \xi, \bar\xi, \rho)= e^{i(\xi^\alpha Q_\alpha +\bar\xi_{\dot\alpha} \bar{Q}^{\dot\alpha})} e^{i\rho R} \, ,
\ee
where $\xi^\alpha, \bar\xi_{\dot\alpha}$ are spacetime independent 2-component Weyl spinors parametrizing the SUSY transformations while the spacetime independent $\rho$ parametrizes the $R$-transforamtion and consider the product $g(0, \xi, \bar\xi, \rho)\Omega(x, \lambda, \bar\lambda, a)$. Exploiting the supersymmetry algebra, Eq. (\ref{SSA}), this product of group elements is seen to again take the form of a product of  translation and coset group elements but with translated spacetime points and coset coordinates. That is, 
\be
 g(0, \xi, \bar\xi, \rho)\Omega(x, \lambda, \bar\lambda, a) =  e^{-ix_{ \mu}^\prime P^\mu} e^{i(\lambda^{\prime\alpha}(x^\prime) Q_\alpha +\bar\lambda^\prime_{\dot\alpha}(x^\prime) \bar{Q}^{\dot\alpha} )}e^{i a^\prime(x^\prime) R}=\Omega (x^\prime, \lambda^\prime, \bar\lambda^\prime, a^\prime)\equiv \Omega^\prime(x^\prime) \, .
\label{gt}
\ee
For any field $\phi^i$ the total ($\Delta$) variation is defined as 
\be
\phi^{i\prime}(x^\prime)=\phi^i(x)+\Delta\phi^i(x) \, .
\ee
The total nonlinear SUSY and $R$ variations of the Nambu-Goldstone degrees of freedom are
\bea
\Delta_Q(\xi,\bar\xi) \lambda^\alpha (x) &=& \xi^\alpha ~~;~~\Delta_R(\rho)\lambda^\alpha (x)= i\rho\lambda^\alpha (x)\cr 
\Delta_Q(\xi,\bar\xi) \bar\lambda_{\dot\alpha} (x) &=& \xi_{\dot\alpha} ~~;~~\Delta_R(\rho)\bar\lambda_{\dot\alpha}(x) = -i\rho\bar\lambda_{\dot\alpha}(x) \cr 
\Delta_Q(\xi,\bar\xi) a(x)&=& 0 ~~~;~~\Delta_R(\rho) a(x) = \rho \, .
\eea
Here $\Delta_Q(\xi,\bar\xi) = \xi^\alpha \Delta_{Q\alpha}+\bar\Delta_{Q\dot\alpha}\bar\xi^{\dot\alpha}$ and $\Delta_R(\rho)=\rho\Delta_R$. 

The accompanying movement in spacetime is given by 
\bea
x^{\prime\mu} &=& x^\mu +\Delta x^\mu \, ,
\eea
with
\be
\Delta_Q(\xi,\bar\xi) x^{\mu} = -\Lambda^\mu(\xi,\bar\xi)~~;~~\Delta_R(\rho) x^\mu =0
\ee
and
\be
\Lambda^\mu (\xi,\bar\xi) = -i[\lambda (x) \sigma^\mu \bar\xi -\xi \sigma^\mu \bar\lambda (x)] \, .
\label{SSV}
\ee

In general, an intrinsic ($\delta$) variation is defined as 
\bea
\delta \phi^i (x) &=& \phi^{i\prime}(x)-\phi^i(x)=\Delta \phi^i (x) - \Delta x^\mu \partial_\mu \phi^i (x) 
\eea
which for nonlinear SUSY and $R$ take the form 
\bea
\delta_Q(\xi,\bar\xi) \lambda^\alpha(x) &=& \xi^\alpha  + \Lambda^\mu (\xi,\bar\xi)\partial_\mu\lambda^\alpha (x) ~~~;~~\delta_R(\rho)\lambda^\alpha (x) =i\rho\lambda^\alpha (x) \cr
\delta_Q(\xi,\bar\xi) \bar\lambda_{\dot\alpha}(x) &=& \bar\xi_{\dot\alpha}  + \Lambda^\mu (\xi,\bar\xi)\partial_\mu \bar\lambda_{\dot\alpha} (x)~~~;~~\delta_R(\rho)\bar\lambda_{\dot\alpha}(x) = -i\rho\bar\lambda_{\dot\alpha}(x) \cr
\delta_Q(\xi,\bar\xi) a(x) &=& \Lambda^\mu(\xi,\bar\xi)\partial_\mu a(x) ~~~~~~~~~~;~~\delta_R(\rho) a(x)=\rho \, ,
\eea
where $\delta_Q(\xi,\bar\xi) =\xi^\alpha\delta_{Q\alpha}+\delta_{Q\dot\alpha}\bar\xi^{\dot\alpha}$ and $\delta_R(\rho)=\rho\delta_R$. A field transforming as $\Lambda^\mu(\xi,\bar\xi)\partial_\mu $ under nonlinear SUSY is said to carry the standard realization\cite{CL2}-\cite{CLLW}. 

In order to construct invariant actions, it proves convenient to introduce covariant derivatives. Towards this end, we define the Mauer-Cartan 1-form as
\be
(\Omega^{-1}d\Omega)(x)=\Omega^{-1}(x,\lambda,\bar\lambda, a) dx^\mu \partial_\mu \Omega (x,\lambda,\bar\lambda, a)
\ee
which, in light of Eq. (\ref{gt}), is invariant under the total variations:
\be
 (\Omega^{-1}d\Omega )^\prime (x^\prime)=(\Omega^{-1}d\Omega)(x) \, .
\ee
Expanding the 1-form in terms of the translation, SUSY and $R$ generators gives
\be
\Omega^{-1}(x,\lambda,\bar\lambda, a)d\Omega (x,\lambda,\bar\lambda, a)=i [-d\omega^\mu (x) P_\mu +d\omega_Q^\alpha (x)Q_\alpha + d\bar\omega_{\bar{Q}~\dot\alpha}(x) \bar{Q}^{\dot\alpha} +d\omega_R(x) R] \, .
\ee

The coefficient coordinate differentials are readily extracted as 
\bea
d\omega^\nu (x)&=& dx^\mu A_\mu~^\nu (x)\cr
d\omega_Q^\alpha (x) &=& dx^\mu e^{-ia(x)} \partial_\mu \lambda^\alpha (x) = d\omega^\mu(x) e^{-ia(x)} D_\mu\lambda^\alpha(x) =d\omega^\mu(x) \nabla_\mu\lambda^\alpha(x)\cr
d\bar\omega_{\bar{Q}~\dot\alpha}(x)&=& e^{ia(x)} \partial_\mu \bar\lambda_{\dot\alpha}(x) =d\omega^\mu(x)e^{ia(x)}D_\mu\bar\lambda_{\dot\alpha}(x)=d\omega^\mu(x)\nabla_\mu\bar\lambda_{\dot\alpha}(x)\cr
d\omega_R (x) &=& dx^\mu \partial_\mu a(x)d\omega^\mu (x)D_\mu a(x)=d\omega^\mu(x)\nabla_\mu a(x)
\, ,
\eea
where\cite{AV}
\be
A_\mu~^\nu = \eta_\mu^\nu + i\lambda\stackrel{\leftrightarrow}{\partial}_\mu\sigma^\nu\bar{\lambda} \, .
\ee

Here we have defined the combined SUSY and $R$ covariant derivatives:
\bea
\nabla_\mu \lambda^\alpha(x) &=&e^{-ia(x)}D_\mu \lambda^\alpha (x)\cr
\nabla_\mu \bar\lambda_{\dot\alpha}(x) &=&e^{ia(x)}D_\mu \bar\lambda_{\dot\alpha}(x)\cr
\nabla_\mu a(x)&=& D_\mu a(x) \, ,
\eea
while 
\be
D_\mu = A^{-1}~_\mu~^\nu\partial_\nu
\ee
is the nonlinear SUSY covariant derivative.

From the invariance of the Mauer-Cartan form, it follows that $d\omega^\mu, \nabla_\mu\lambda, \nabla_\mu\bar\lambda, \nabla_\mu a$ are all invariant under the total $\Delta$-variations
\bea
(d\omega)^{\prime~\mu}(x^\prime) &=& d\omega^\mu (x) \cr
(\nabla_\mu \lambda)^{\prime~\alpha}(x^\prime)&=&\nabla_\mu \lambda^\alpha(x)\cr
(\nabla_\mu \bar\lambda)^\prime_{\dot\alpha}(x^\prime)&=&\nabla_\mu \bar\lambda_{\dot\alpha}(x)\cr
(\nabla_\mu a)^\prime(x^\prime)&=&\nabla_\mu a(x) \, .
\eea
Under the intrinsic nonlinear SUSY variations, $\delta_Q(\xi,\bar\xi)$, the covariant derivatives, $\nabla_\mu\phi^i$,  transform as the standard realization, while being invariant under the intrinsic $\delta_R(\rho)$ variation:
\be
\delta_Q(\xi,\bar\xi)(\nabla_\mu\phi^i)(x) = \Lambda^\nu(\xi,\bar\xi)\partial_\nu(\nabla_\mu\phi^i)(x)~~;~~\delta_R(\rho)(\nabla_\mu\phi^i)(x)=0 \, .
\ee
The SUSY covariant derivative, $D_\mu\phi^i$, also transform as the standard realization under the intrinsic $\delta$-variations:\\
\be
\delta_Q(\xi,\bar\xi)\nabla_\mu\phi^i(x)=\Lambda^\nu(\xi,\bar\xi)\partial_\nu(\nabla_\mu \phi^i(x)) \, .
\ee

Using the $\Delta$ invariance of $d\omega^\mu$ along with 
\be
dx^{\prime ~\mu} =\frac{\partial x^{\prime~\mu}}{\partial x^\nu}dx^\nu = dx^\nu (\eta_\nu^\mu -\partial_\nu \Lambda^\mu)\equiv dx^\nu G_{\nu}~^\mu \, ,
\ee
it follows that 
\bea
A^\prime~_\mu ~^\nu (x^\prime) &=& G^{-1}~_\mu~^\rho (x)A_\rho~^\nu (x)\cr
A^{-1~\prime}~_\nu ~^\mu (x^\prime) &=& A^{-1}~_\nu~^\rho (x)G~_\rho~^\mu (x) 
\eea
producing the intrinsic variations:
\bea
\delta_Q(\xi,\bar\xi) A_\mu~^\nu  &=& \Delta_Q(\xi,\bar\xi) A_\mu~^\nu  -\Delta_Q(\xi,\bar\xi) x^\rho\partial_\rho A_\mu~^\nu\cr
&=& (\partial_\mu \Lambda^\rho (\xi,\bar\xi))A_\rho~^\nu  +\Lambda^\rho(\xi,\bar\xi)\partial_\rho A_\mu~^\nu \cr
\delta_Q(\xi,\bar\xi) A^{-1}~_\nu~^\mu &=& \Delta_Q(\xi,\bar\xi) A^{-1}~_\nu~^\mu (x) -\Delta x^\rho \partial_\rho A^{-1}~_\nu~^\mu \cr
 &=& -A^{-1}~_\nu~^\rho \partial_\rho\Lambda^\mu (\xi,\bar\xi)+\Lambda^\rho(\xi,\bar\xi)\partial_\rho A^{-1}~_\nu~^\mu \cr
\delta_Q(\xi,\bar\xi)\det{A}~&=&\det{A}~A^{-1}_\nu~^\mu \delta_Q(\xi,\bar\xi)A_\mu~^\nu \cr
&=& \partial_\mu [\Lambda^\mu(\xi,\bar\xi)~\det{A}]
\eea
and
\bea
\delta_R(\rho) A_\mu~^\nu &=&0 \, .
\eea

To construct invariant actions, note that the product $d^4x \det{A}$ is $\Delta$ invariant, 
\be
d^4x^\prime \det{A^\prime}(x^\prime)=d^4x \det{A}(x) \, ,
\ee
and thus $A_\mu~^\nu$ can be viewed as a ``vierbein''. It follows that the action
\be
I=\int d^4x {\cal L}(x)
\ee
with Lagrangian
\be
{\cal L} = \det{A}~{\cal O}(\nabla a,\nabla\lambda,\nabla\bar\lambda)
\ee
and ${\cal O}$ is any Lorentz singlet function is invariant under nonlinear SUSY as well as $R$-transformations. The leading terms in a derivative expansion of the Lagrangian are
\be
{\cal L} = -\frac{f_s^4}{2}\det{A}~-\frac{f_a^2}{2}\det{A}~D_\mu a D^\mu a \, .
\ee
Here $f_s$ is the SUSY breaking scale and $f_a$ is the $R$-symmetry breaking scale. 
In general, these scales are independent. In the absence of the spontaneous $R$ symmetry breaking and hence the $R$ axion, this reduces to the Akulov-Volkov action\cite{AV},\cite{WS}. One can also include a nonlinearly SUSY invariant but soft $R$-symmetry breaking mass term by modifying the Lagrangian to be of the form
\be
{\cal L} = -\frac{f_s^4}{2}\det{A}~-\frac{f_a^2}{2}\det{A}~D_\mu a D^\mu a -\frac{1}{2}m_a^2 f_a^2 \det{A}~ a^2 \ .
\label{L1}
\ee

Note that ${\cal L}$ can alternatively be written by defining a ``metric'' 
\be
g^{\mu\nu}= A^{-1}~_\rho~^\mu \eta^{\rho\sigma}A^{-1}~_\sigma~^\nu  = g^{\nu\mu}
\ee
as 
\be
{\cal L} = -\frac{f_s^4}{2}\sqrt{\det{(-g)}}~-\frac{f_a^2}{2}\sqrt{\det{(-g)}}~\partial_\mu a g^{\mu\nu} \partial_\nu a -\frac{1}{2}m_a^2 f_a^2 \sqrt{\det{(-g)}} ~a^2 \, .
\ee
Here $\eta^{\mu\nu}$ is the Minkowski space metric with signature $(-1,1,1,1)$.

\newpage

\newsection{Symmetry currents}

For any variations  $\Delta x^\mu, \delta\phi^i$ (not just SUSY and R), one has
\be
\delta{\cal L} = -\partial_\mu J_N^\mu +\sum_i\delta\phi^i\frac{\delta I}{\delta\phi^i}
\ee
where 
\be
J_N^\mu =- \sum_i\delta\phi^i \frac{\partial {\cal L}}{\partial \partial_\mu \phi^i}
\ee
and the Euler-Lagrange derivative is
\be
\frac{\delta I}{\delta\phi^i}= \frac{\partial {\cal L}}{\partial \phi^i} -\partial_\mu \frac{\partial {\cal L}}{\partial \partial_\mu \phi^i} \, .
\ee

Using
\be
\delta {\cal L} =\Delta {\cal L} -\Delta x^\mu \partial_\mu {\cal L} = -\partial_\mu[\Delta x^\mu {\cal L}] +\Delta {\cal L} +(\partial_\mu \Delta x^\mu){\cal L}
\ee
and defining the Noether current 
\be
J^\mu = J_N^\mu -\Delta x^\mu {\cal L}
\ee
gives Noether's theorem
\be
\partial_\mu J^\mu = -\Delta{\cal L}-(\partial_\mu \Delta x^\mu) {\cal L} + \sum_i \delta \phi^i \frac{\delta I}{\delta \phi^i}\, .
\ee
The general current form can be rewriten using the canonical energy-momentum tensor 
\be
T^\mu~_\nu =-\sum_i \partial_\nu \phi^i \frac{\partial {\cal L}}{\partial\partial_\mu \phi^i}+\eta^\mu_\nu {\cal L}
\ee
as
\be
J^\mu =-\sum_i \Delta \phi^i \frac{\partial {\cal L}}{\partial \partial_\mu \phi^i} -\Delta x^\nu T^\mu~_\nu \, .
\ee

For the model described by the Lagrangian of Eq. (\ref{L1}), the conserved canonical energy-momentum tensor is simply 
\be
T^\mu~_\nu = 
A^{-1}~_\nu~^\mu {\cal L} +f_a^2 \det{A}~ D^\rho a A^{-1}_\rho~^\mu  D_\nu a
\ee
and satisfies
\be
\partial_\mu T^\mu~_\nu = \sum_i \partial_\nu\phi^i \frac{\delta I}{\delta\phi^i} \, .
\ee
Note that the energy-momentum tensor starts as a positive cosmological constant associated with the spontaneous SUSY breaking
\be
<0|T^{00}|0> =  \frac{f_s^4}{2} \, .
\ee

Using the Noether construction, it is straightforward to obtain the form of the supersymmetry and $R$ currents along with their (non-) conservation laws\cite{CL1}. The conserved supersymmetry currents are 
\bea
Q^\mu(\xi,\bar\xi) &=&\xi^\alpha Q^\mu_\alpha +\bar{Q}^\mu_{\dot\alpha}\bar\xi^{\dot\alpha}\cr
&=&2T^\mu~_\nu \Lambda^\nu (\xi,\bar\xi) \, ,
\eea
where
\be
Q^\mu_\alpha =2i T^\mu~_\nu (\sigma^\nu \bar\lambda)_\alpha ~~;~~\bar{Q}^\mu_{\dot\alpha}=-2i T^\mu~_\nu (\lambda\sigma^\nu)_{\dot\alpha}
\ee
and satisfy
\be
\partial_\mu Q^\mu(\xi,\bar\xi) = \sum_i \delta_Q(\xi,\bar\xi)\phi^i \frac{\delta I}{\delta\phi^i} \, .
\ee

For the R-symmetry current, one extracts the current 
\be
R^\mu = f_a^2 \det{A}~ A^{-1}_\rho~^\mu D^\rho a - 2 T^\mu~_\nu (\lambda \sigma^\nu \bar\lambda)
\ee
whose divergence
\be
\partial_\mu R^\mu = m_a^2 f_a^2 \det{A}~ a +\frac{\delta I}{\delta a} +i(\lambda^\alpha \frac{\delta I}{\delta \lambda^\alpha}+\frac{\partial I}{\partial \bar\lambda^{\dot\alpha}}\bar\lambda^{\dot\alpha})
\ee
displays the soft $R$ symmetry breaking.

Note that 
\bea
R^\mu &=& f_a \partial^\mu a +... \nonumber \\
Q^\mu_\alpha &=& i f_s^2(\sigma^\mu \bar\lambda)_\alpha +... \nonumber \\ 
\bar{Q}^\mu_{\dot\alpha} &=& -if_s^2 (\lambda \sigma^\mu )_{\dot\alpha)}+... \nonumber \\
T^{\mu\nu} &=& -\eta^{\mu\nu}\frac{f_s^4}{2}+...
\eea
so that $R^\mu$ interpolates for the $R$-axion field $a$, while the supersymmetry currents interpolate for the Goldstino fields $\lambda_\alpha, \bar\lambda_{\dot\alpha}$.

In addition, ${\cal L}$  also possesses a softly broken (by the $R$-axion mass term) shift symmetry\cite{SL}  defined by 
\bea
\delta (\zeta)a &=& \zeta \cr
\delta (\zeta)\lambda^\alpha &=& 0 \cr
\delta (\zeta)\bar\lambda^{\dot\alpha} &=& 0 \, . 
\eea
The associated current and non-conservation equation are given by 
\be
j^\mu = f_a^2 \det{A}~ A^{-1}_\rho~^\mu D^\rho a
\ee
and
\be
\partial_\mu j^\mu = m_a^2 f_a^2 \det{A}~ a + \frac{\delta S}{\delta a} \, .
\ee

For the case of linearly realized supersymmetry, the various currents are components of a supercurrent\cite{FZ}-\cite{CLSC} and are related via SUSY transformations. Under nonlinear supersymmetry, the  currents transform as 
\bea
\delta_Q(\xi,\bar\xi) R^\mu &=&i(\xi^\alpha Q^\mu_\alpha  -\bar{Q}^\mu_{\dot\alpha}\bar\xi^{\dot\alpha}) +\partial_\rho [\Lambda^\rho (\xi,\bar\xi) R^\mu -
\Lambda^\mu (\xi,\bar\xi) R^\rho] + \Lambda^\mu (\xi,\bar\xi) \partial_\rho R^\rho \nonumber \\
\delta_Q(\xi,\bar\xi)Q^\mu_\alpha &=& 2i (\sigma^\nu \bar\xi )_\alpha T^\mu~_\nu +\partial_\nu [\Lambda^\nu(\xi,\bar\xi)Q^\mu_\alpha -\Lambda^\mu(\xi,\bar\xi)Q^\nu_\alpha ] +\Lambda^\mu(\xi,\bar\xi)\partial_\nu Q^\nu_\alpha \nonumber \\
\delta_Q(\xi,\bar\xi)\bar{Q}^\mu_{\dot\alpha} &=& -2i (\xi\sigma^\nu )_{\dot\alpha} T^\mu~_\nu +\partial_\nu [\Lambda^\nu(\xi,\bar\xi)\bar{Q}^\mu_{\dot\alpha} -\Lambda^\mu(\xi,\bar\xi)\bar{Q}^\nu_{\dot\alpha} ] +\Lambda^\mu(\xi,\bar\xi)\partial_\nu \bar{Q}^\nu_{\dot\alpha} \nonumber \\
\delta_Q(\xi,\bar\xi) T^\mu~_\nu &=& \partial_\rho [\Lambda^\rho(\xi,\bar\xi)T^\mu~_\nu-\Lambda^\mu(\xi,\bar\xi)T^\rho~_\nu]+\Lambda^\mu(\xi,\bar\xi)\partial_\rho T^\rho~_\nu \, .
\eea
Note that the terms 
\bea
&&\partial_\rho [\Lambda^\rho (\xi,\bar\xi) R^\mu -
\Lambda^\mu (\xi,\bar\xi) R^\rho] \nonumber \\
&&\partial_\nu [Q^\mu_\alpha \Lambda^\nu(\xi,\bar\xi)-Q^\nu_\alpha \Lambda^\mu(\xi,\bar\xi)]\nonumber \\
&& \partial_\nu [\bar{Q}^\mu_{\dot\alpha} \Lambda^\nu(\xi,\bar\xi)-\bar{Q}^\nu_{\dot\alpha} \Lambda^\mu(\xi,\bar\xi)]\nonumber \\
&& \partial_\rho [ \Lambda^\rho(\xi,\bar\xi)T^{\mu\nu}-\Lambda^\mu(\xi,\bar\xi)T^{\rho\nu}]
\eea
are simply Belinfante improvements each being algebraically divergenceless. They can absorbed into defining improved currents.

Thus, under the nonlinear SUSY, $R^\mu$ transforms into  $Q^\mu_{\alpha}, \bar{Q}^\mu_{\dot\alpha}$ and $Q^\mu_{\alpha}, \bar{Q}^\mu_{\dot\alpha}$ transforms into $T^\mu~_\nu$. This holds even though \\
(i) $a$ and $\lambda, \bar\lambda$ need not be SUSY partners in the underlying theory \\
(ii) the dynamics responsible for SUSY breaking and $R$ breaking may have different origins \\ 
(iii) $f_s$ need not equal $f_a$. \\
Since they are related by SUSY transformations, there will be relations among  $R$ and SUSY current correlators. Moreover, since these currents 
interpolate for the $R$-axion and Goldstinos, the current correlator relations will translate into relations among Green functions (and S-matrix elements) containing $R$-axions and Goldstinos.

In closing this section, note that the softly broken $R$-axion shift symmetry current, $j^\mu$, has the nonlinear SUSY transformation law
\be
\delta_Q(\xi,\bar\xi) j^\mu =\partial_\rho [\Lambda^\rho(\xi,\bar\xi) j^\mu - \Lambda^\mu(\xi,\bar\xi) j^\rho ] +\Lambda^\mu(\xi,\bar\xi) \partial_\rho j^\rho 
\ee
which is just a sum of the Belinfante improvement term and the divergence of $j^\mu$ itself.

\newpage

\newsection{Superconformal variations and relations among symmetry breakings }

For linear realizations of supersymmetry, not only are the $R$ current, supersymmetry currents and the energy momentum tensor related by supersymmetry transformations, but the explicit breakings of the $R$ and dilatation symmetries are related via SUSY transformations to the breaking of the superconformal symmetry. In this section, we examine the analog of this connection for nonlinear SUSY. 

Our first step is to construct the intrinsic variations\cite{sconformal} of the $R$ axion and Goldstino fields which satisfy the (graded) superconformal algebra\cite{WZ},\cite{F}
\bea
&&[\delta_D,\delta_{M\mu\nu}]= 0 ~~;~~[\delta_D,\delta_{P\mu}]= \delta_{P\mu}~~;~~[\delta_D,\delta_{K\mu}]= -\delta_{K\mu}\cr
&&[\delta_{M\mu\nu},\delta_{K\rho}]= \eta_{\mu\rho}\delta_{K\nu}-\eta_{\nu\rho}\delta_{K\mu}~~;~~[\delta_{P\mu},\delta_{K\nu}]= 2(\eta_{\mu\nu}\delta_D -\delta_{M\mu\nu})~~;~~[\delta_{P\mu},\delta_{K\nu}]= 0 \cr
&&[\delta_{K\mu},\delta_R]= 0 ~~;~~[\delta_{K\mu}, \delta_{K\nu}]=0~~;~~[\delta_D,\delta_R]= 0  \cr
&&[\delta_D,\delta_{Q\alpha}]= \frac{1}{2}\delta_{Q\alpha}~~;~~[\delta_D,\bar\delta_{Q\dot\alpha}]= \frac{1}{2}\bar\delta_{Q\dot\alpha}\cr
&&[\delta_{K\mu},\delta_{Q\alpha}]=\sigma_{\mu\alpha\dot\alpha}\bar\delta_S^{\dot\alpha} ~~;~~[\delta_{K\mu},\bar\delta_{Q\dot\alpha}]=\delta_S^{\alpha}\sigma_{\mu\alpha\dot\alpha} \cr
&&[\delta_{M\mu\nu},\delta_{S\alpha}]= -i\frac{1}{2}\sigma_{\mu\nu_\alpha}~^\beta \delta_{S\beta}~~;~~ 
[\delta_{M\mu\nu},\bar\delta_{S\dot\alpha}]= -i\frac{1}{2}\bar\sigma_{\mu\nu\dot\alpha\dot\beta} \bar\delta_{S}^{\dot\beta} \cr
&&[\delta_{P\mu},\delta_{S\alpha}]=\sigma_{\mu\alpha\dot\alpha}\bar\delta_Q^{\dot\alpha} ~~;~~
[\delta_{P\mu},\bar\delta_{S\dot\alpha}]=\delta_Q^\alpha\sigma_{\mu\alpha\dot\alpha}\cr
&&[\delta_R,\delta_{S\alpha}]= i\delta_{S\alpha} ~~;~~
[\delta_R,\bar\delta_{S\dot\alpha}] = -\bar\delta_{S\dot\alpha} \cr
&&[\delta_D,\delta_{S\alpha}] = -\frac{1}{2}\delta_{S\alpha} ~~;~~
[\delta_D,\bar\delta_{S\dot\alpha}] = -\frac{1}{2}\bar\delta_{S\dot\alpha} \cr
&&[\delta_{K\mu},\delta_{S\alpha}]= 0 ~~;~~
[\delta_{K\mu},\bar\delta_{S\dot\alpha}]= 0 ~~;~~
\cr
&&\{\delta_{S\alpha},\bar\delta_{S\dot\alpha}\}= -2i\sigma^\mu_{\alpha\dot\alpha}\delta_{K\mu} \cr
&&\{\delta_{Q\alpha},\delta_{S\beta}\}=-(\sigma^{\mu\nu}_{\alpha\beta}\delta_{\mu\nu}+2i\epsilon_{\alpha\beta}\delta_D +3\epsilon_{\alpha\beta}\delta_R) \cr
&&\{\delta_{Q\alpha},\delta_{S\beta}\}=\bar\sigma^{\mu\nu}_{\dot\alpha\dot\beta}\delta_{\mu\nu}-2i\epsilon_{\dot\alpha\dot\beta}\delta_D +3\epsilon_{\dot\alpha\dot\beta}\delta_R \cr
&&\{\delta_{Q\alpha},\bar\delta_{S\dot\alpha}\}= 0 ~~;~~
\{\delta_{S\alpha},\bar\delta_{Q\dot\alpha}\}= 0 ~~;~~
\{\delta_{S\alpha},\delta_{S\alpha}\}= 0 ~~;~~
\{\bar\delta_{S\dot\alpha},\bar\delta_{S\dot\alpha}\}= 0 
\eea
where $\delta_{P\mu}=\partial_\mu, \delta_{M\mu\nu}, \delta_{K\mu}$ are the translation, angular momentum and special conformal variations respectively and $\delta_{S\alpha}, \bar\delta_{S\dot\alpha}$ are the superconformal variations. Note that the angular momentum variations also satisfy 
\bea
&&[\delta_{M\mu\nu}, \delta_{M\lambda\rho}]= \eta_{\mu\lambda}\delta_{M\nu\rho}-\eta_{\mu\rho}\delta_{M\nu\lambda}-\eta_{\nu\lambda}\delta_{M\mu\rho}+\eta_{\nu\rho}\delta_{M\mu\lambda}\cr
&&[\delta_{M\mu\nu},\delta_{P\rho}]=\eta_{\mu\rho}\delta_{P\nu}-\eta_{\nu\rho}\delta_{P\mu}~~;~~
[\delta_{M\mu\nu},\delta_R]=0 \cr 
&&[\delta_{M\mu\nu},\delta_{Q\alpha}]=-i\frac{1}{2}\sigma^{\mu\nu}_\alpha~^\beta \delta_{Q\beta}~~;~~
[\delta_{M\mu\nu},\bar\delta_{Q\dot\alpha}]=-i\frac{1}{2}\bar\sigma^{\mu\nu}_{\dot\alpha\dot\beta} \bar\delta_Q^{\dot\beta}
\eea

After some straightforward, but rather tedious algebra, the (intrinsic) superconformal  variations of the Nambu-Goldstone fields are extracted as
\bea
\delta_{S\alpha} a &=& 3\lambda_\alpha - \left(i(\sigma^\nu \bar\sigma^\mu \lambda)_\alpha x_\nu -2\lambda_\alpha (\lambda\sigma^\mu\bar\lambda)\right)\partial_\mu a \cr
\delta_{S\alpha} \lambda^\beta &=& 4i\lambda^\alpha\lambda^\beta   - \left(i(\sigma^\nu \bar\sigma^\mu \lambda)_\alpha x_\nu -2\lambda_\alpha (\lambda\sigma^\mu\bar\lambda)\right)\partial_\mu \lambda^\beta \cr
\delta_{S\alpha}  \bar\lambda^{\dot\beta}&=& -x^\nu\bar\sigma_\nu^{\dot\beta\alpha} -2i \lambda^\alpha\bar\lambda^{\dot\beta} - \left(i(\sigma^\nu \bar\sigma^\mu \lambda)_\alpha x_\nu -2\lambda_\alpha (\lambda\sigma^\mu\bar\lambda)\right)\partial_\mu \bar\lambda^{\dot\beta}
\label{SCV}
\eea

\bea
\bar\delta_{S\dot\alpha} a &=& 3\bar\lambda^{\dot\alpha} - \left(-i (\bar\lambda \bar\sigma^\mu\sigma^\nu)_{\dot\alpha}x_\nu -2\bar\lambda_{\dot\alpha}(\lambda\sigma\bar\lambda) \right)\partial_\mu a \cr
\bar\delta_{S\dot\alpha} \lambda^\beta &=& x^\nu\bar\sigma_\nu^{\dot\alpha \beta} +2i\bar\lambda^{\dot\alpha}\lambda^{\beta} - \left(-i (\bar\lambda \bar\sigma^\mu\sigma^\nu)_{\dot\alpha}x_\nu -2\bar\lambda_{\dot\alpha}(\lambda\sigma^\mu\bar\lambda) \right)\partial_\mu \lambda^\beta \cr
\bar\delta_{S\dot\alpha} \bar\lambda^{\dot\beta}&=& -4i \bar\lambda^{\dot\alpha}\bar\lambda^{\dot\beta} - \left(-i (\bar\lambda \bar\sigma^\mu\sigma^\nu)_{\dot\alpha}x_\nu -2\bar\lambda_{\dot\alpha}(\lambda\sigma^\mu\bar\lambda) \right)\partial_\mu \bar\lambda^{\dot\beta}
\label{SCV1}
\eea 

Notice that the superconformal symmetries are also spontaneously broken, but the associated Nambu-Goldstone fermions, $\lambda_S^\alpha~,~\bar\lambda_S^{\dot\alpha}$ are not independent degrees of freedom. Rather, one finds that
\bea
\lambda_S^\alpha &=& -\frac{1}{4} (\sigma^\mu\partial_\mu\bar\lambda)^\alpha +... \cr
\bar\lambda_S^{\dot\alpha} &=& -\frac{1}{4} (\partial_\mu\lambda\sigma^\mu)^{\dot\alpha} +...
\eea
so that
\bea
\delta_{S\alpha}\lambda_S^\beta = \delta_\alpha^\beta +... ~~&;&~~\bar\delta_{S\dot\alpha}\bar\lambda_S^{\dot\beta} = -\delta_{\dot\alpha}^{\dot\beta} +... \, .
\eea
The fact that there can be spontaneously broken spacetime symmetries without independent Nambu-Goldstone fields is  not in conflict with Goldstone's theorem\cite{G} which guarentees an independent Nambu-Goldstone field for every spontaneously broken global symmetry\cite{inversehiggs}-\cite{CLD}.

For the dilatation variations, one finds 
\bea
\delta_D a&=&  x^\mu\partial_\mu a \cr
\delta_D \lambda^\alpha &=&(-\frac{1}{2}+x^\mu\partial_\mu)\lambda^\alpha\cr
\delta_D \bar\lambda^{\dot\alpha} &=&(-\frac{1}{2}+x^\mu\partial_\mu)\bar\lambda^{\dot\alpha} \, ,
\eea
which corresponds to a linear representation.

Recalling that the intrinsic and total variations are related as 
\be
\delta \phi^i = \Delta \phi^i -\Delta x^\mu \partial_\mu \phi^i \, ,
\ee
we extract the total ($\Delta$) scale variations corresponding to the spacetime scaling \\
$\Delta_D x^\mu = - x^\mu $ as 
\bea
\Delta_D a &=& 0~~;~~ \Delta_D \lambda^\alpha = -\frac{1}{2}\lambda^\alpha ~~;~~\Delta_D \bar\lambda_{\dot\alpha} = -\frac{1}{2}\bar\lambda_{\dot\alpha} 
\eea
which fixes the scaling weights as
\bea
&&d_a = 0 ~~;~~ d_\lambda = d_{\bar\lambda}=-\frac{1}{2} \, .
\eea
It follows that
\bea
\Delta_D {\cal L}&=& - f_a^2 \det{A}~ D^\mu a D_\mu a \, .
\eea
In fact, the Nambu-Goldstone particle  associated with any spontaneously broken symmetry which commutes with the dilatation charge is constrained to have scaling weight zero\cite{CLL}. With the variations in hand, the dilatation current is readily constructed as 
\bea
{\cal D}^\mu &=& x^\nu T^\mu~_\nu \, .
\eea
Its divergence 
\bea
\partial_\mu {\cal D}^\mu &=& T^\mu~_\mu +x^\nu \partial_\mu T^\mu~_\nu \cr
&=& A_\mu~^\nu T^\mu~_\nu +x^\nu \partial_\mu T^\mu~_\nu -\frac{1}{2}(\lambda^\alpha\frac{\delta I}{\delta \lambda^\alpha}-\frac{\delta I}{\delta \bar\lambda^{\dot\alpha}}\bar\lambda^{\dot\alpha})\cr
&=&[f_s\frac{\partial}{\partial f_s}+f_a\frac{\partial}{\partial f_a}+m_a\frac{\partial}{\partial m_a}]{\cal L}+\sum_i \delta_D\phi^i\frac{\delta I}{\delta \phi^i}
\eea
exhibits the explicit scale symmetry breaking. Note that there are independent breakings arising from the spontaneous SUSY breaking scale, $f_s$, the spontaneous $R$ symmetry breaking scale, $f_a$, and the soft $R$ symmetry breaking mass term, $m_a$. Under SUSY, the dilatation current transforms as 
\bea
\delta_Q(\xi,\bar\xi){\cal D}^\mu &=& -\frac{1}{2}Q^\mu (\xi,\bar\xi)+\partial_\rho [\Lambda^\rho (\xi,\bar\xi) {\cal D}^\mu -\Lambda^\mu (\xi,\bar\xi){\cal D}^\rho ] \cr
&&~~~~+\Lambda^\mu (\xi,\bar\xi)\partial_\rho {\cal D}^\rho  \, .
\eea

The special conformal intrinsic variations take the form 
\bea
\delta_{K\mu} a &=& -3\lambda\sigma_\mu\bar\lambda  +\left(\eta_\mu^\nu x^2 -2 x^\nu x_\mu +\eta^\nu_\mu(\lambda\lambda) (\bar\lambda\bar\lambda)\right)\partial_\nu a \cr
\delta_{K\mu} \lambda^\alpha &=& (x_\mu -2i\lambda\sigma_\mu\bar\lambda)\lambda^\alpha -ix^\nu(\sigma_{\mu\nu}\lambda)^\alpha+\left(\eta_\mu^\nu x^2 -2 x^\nu x_\mu +\eta^\nu_\mu(\lambda\lambda) (\bar\lambda\bar\lambda)\right)\partial_\nu \lambda^\alpha \cr
\delta_{K\mu} \bar\lambda^{\dot\alpha} &=& (x_\mu -2i\lambda\sigma_\mu\bar\lambda)\bar\lambda^{\dot\alpha} -ix^\nu(\bar\sigma_{\mu\nu}\bar\lambda)^{\dot\alpha}+\left(\eta_\mu^\nu x^2 -2 x^\nu x_\mu +\eta^\nu_\mu(\lambda\lambda) (\bar\lambda\bar\lambda)\right)\partial_\nu \bar\lambda^{\dot\alpha} 
\eea
while the associated special conformal total variations of the fields are 
\bea
\Delta_{K\mu} a &=& -3\lambda\sigma_\mu\bar\lambda \cr
\Delta_{K\mu}\lambda^\alpha &=& (x_\mu -2i\lambda\sigma_\mu\bar\lambda)\lambda^\alpha -ix^\nu(\sigma_{\mu\nu}\lambda)^\alpha \cr
\Delta_{K\mu}\bar\lambda^{\dot\alpha}&=& (x_\mu -2i\lambda\sigma_\mu\bar\lambda)\bar\lambda^{\dot\alpha} -ix^\nu(\bar\sigma_{\mu\nu}\bar\lambda)^{\dot\alpha} 
\eea
while the spacetime point varies as
\bea
\Delta_{K\mu} x^\nu &=& -(\eta^\nu_\mu x^2-2x_\mu x^\nu)-\eta^\nu_\mu(\lambda\lambda) (\bar\lambda\bar\lambda)
\eea
Note that realizing the supersymmetry nonlinearly requires that the special conformal tranformations be nonlinearly realized. This is reminiscent of the situation which occurs in the  coupling of gauge fields with nonlinearly realized SUSY. In that case, if one demands that the gauge field transforms as a nonlinear SUSY standard realization, then its gauge transformation is nonstandard\cite{CLLW}. Note, however, that neither the special conformal nor the dilatation symmetries are spontaneously broken in this realization.

From the intrinsic superconformal variations (c.f. Eqs. (\ref {SCV})-(\ref{SCV1})), the total ($\Delta$) superconformal variations are immediately gleaned as 
\bea
\Delta_{S\alpha} a &=& 3\lambda_\alpha ~~;~~ \bar\Delta_{S\dot\alpha}a= 3\bar\lambda^{\dot\alpha}\cr
\Delta_{S\alpha} \lambda^\beta &=& 4i\lambda^\alpha\lambda^\beta  ~~;~~ \bar\Delta_{S\dot\alpha}\lambda^\beta = x^\nu\bar\sigma_\nu^{\dot\alpha \beta} +2i\bar\lambda^{\dot\alpha}\lambda^{\beta}\cr
\Delta_{S\alpha}  \bar\lambda^{\dot\beta}&=& -x^\nu\bar\sigma_\nu^{\dot\beta\alpha} -2i \lambda^\alpha\bar\lambda^{\dot\beta} ~~;~~ \bar\Delta_{S\dot\alpha}\bar\lambda^{\dot\beta}= -4i \bar\lambda^{\dot\alpha}\bar\lambda^{\dot\beta}
\eea
with the superconformal variations of the spacetime point given by
\bea
\Delta_{S\alpha} x^\mu &=& i(\sigma^\nu \bar\sigma^\mu \lambda)_\alpha x_\nu -2\lambda_\alpha (\lambda\sigma^\mu\bar\lambda) \cr
\bar\Delta_{S\dot\alpha} x^\mu &=& -i (\bar\lambda \bar\sigma^\mu\sigma^\nu)_{\dot\alpha} x_\nu -2\bar\lambda_{\dot\alpha}(\lambda\sigma\bar\lambda)
\eea

The superconformal currents are now readily computed as 
\bea 
S^\mu(\eta,\bar\eta)&=& \eta^\alpha S^\mu_\alpha +\bar{S}^\mu_{\dot\alpha} \bar\eta^{\dot\alpha} \cr 
&=&  Q^\mu (\eta\bar\sigma^\rho x_\rho,\bar\sigma^\rho\eta x_\rho)+(\eta\lambda+\bar\eta\bar\lambda)(2R^\mu +j^\mu) \cr
&=&  [(\bar\eta \bar\sigma^\nu)^\alpha Q_{\mu \alpha } +\bar{Q}_{\mu \dot\alpha}(\bar\sigma^\nu\eta)^{\dot\alpha}]x_\nu +(\eta\lambda+\bar\eta\bar\lambda)(2R^\mu +j^\mu) \, .
\eea

Under nonlinear SUSY, the superconformal currents transform as 
\bea
\delta_Q(\xi,\bar\xi) S^\mu (\eta,\bar\eta) &=&3(\eta\xi+\bar\eta\bar\xi)R^\mu +2i(\eta\xi-\bar\eta\bar\xi){\cal D}^\mu \cr
&& -(\xi\sigma^{\nu\rho}\eta + \bar\eta \bar\sigma^{\nu\rho}\bar\xi)M^\mu~_{\nu\rho} +\partial_\rho [\Lambda^\rho (\xi,\bar\xi) S^\mu(\eta,\bar\eta)\cr
&& -\Lambda^\mu (\xi,\bar\xi)S^\rho(\eta,\bar\eta)] +\Lambda^\mu (\xi,\bar\xi)\partial_\nu S^\nu (\eta,\bar\eta) \, ,
\label{SCC}
\eea
where the conserved angular momentum tensor is 
\bea
M^{\mu\nu\rho}&=& T^{\mu\nu} x^\rho - T^{\mu\rho} x^\nu -\frac{1}{2}[\lambda \sigma^\lambda\bar\sigma^{\nu\rho}\bar\lambda +\lambda\sigma^{\nu\rho}\sigma^\lambda\bar\lambda]T^\mu~_\lambda \nonumber
\eea
and the angular momentum variations are
\bea
\delta_{M\mu\nu}a&=& (x_\mu\partial_\nu-x_\nu\partial_\mu )a \cr
\delta_{M\mu\nu}\lambda^\alpha &=&(x_\mu\partial_\nu-x_\nu\partial_\mu)\lambda^\alpha+\frac{i}{2}(\sigma_{\mu\nu}\lambda)^\alpha \cr
\delta_{M\mu\nu}\bar\lambda^{\dot\alpha} &=&(x_\mu\partial_\nu-x_\nu\partial_\mu)\bar\lambda^{\dot\alpha}+\frac{i}{2}(\bar\sigma_{\mu\nu}\bar\lambda)^{\dot\alpha} \, .
\eea

Taking the divergence of Eq. (\ref{SCC}) yields
\bea
\delta_Q(\xi,\bar\xi) \partial_\mu S^\mu (\eta,\bar\eta) &=&3(\eta\xi+\bar\eta\bar\xi)\partial_\mu R^\mu +2i(\eta\xi-\bar\eta\bar\xi)\partial_\mu{\cal D}^\mu \cr
&& -(\xi\sigma^{\nu\rho}\eta + \bar\eta \bar\sigma^{\nu\rho}\bar\xi)\partial_\mu M^\mu~_{\nu\rho}  +\partial_\mu[\Lambda^\mu (\xi,\bar\xi)\partial_\nu S^\nu (\eta,\bar\eta)]\, .
\eea
Thus the divergence of the $R$-symmetry current and the divergence of the dilatation current are tied to the divergence of the superconformal current through a nonlinear SUSY transformation. Since the divergences of these currents give the explicit breakings of these symmetries in the action, these breakings are also related via the nonlinear SUSY transformation.

$\hspace{.3in}$

\noindent
This work was supported in part by the U.S. Department of Energy 
under grant DE-FG02-91ER40681A29 (Task B).
\pagebreak

\end{document}